\documentclass[twocolumn,twocolappendix]{aastex63}

\usepackage{newtxtext,newtxmath}
\usepackage{bbm}

\usepackage[T1]{fontenc}
\usepackage{ae,aecompl}
\usepackage{comment}

\usepackage{graphicx,tabularx}	
\usepackage{amsmath}	
\usepackage{amssymb}	
\usepackage{booktabs}
\usepackage[inline,shortlabels]{enumitem}
\setlist{itemjoin* = { and\enspace}}
\usepackage{color}

\defcitealias{2021arXiv210503344Z}{Z22}
\received{XXX}
\revised{XXX}
\accepted{XXX}
\submitjournal{ApJ}

\shorttitle{Reionization parameter inference}
\shortauthors{Zhao, Mao \& Wandelt}
\graphicspath{{./}}

\begin{document}

\title{Implicit Likelihood Inference of Reionization Parameters from the 21 cm Power Spectrum}

\correspondingauthor{Xiaosheng Zhao, Yi Mao}
\email{zhaoxs18@mails.tsinghua.edu.cn (XZ), ymao@tsinghua.edu.cn (YM)}

\author[0000-0002-8328-1447]{Xiaosheng Zhao}
\affiliation{Department of Astronomy, Tsinghua University, Beijing 100084, China}

\author[0000-0002-1301-3893]{Yi Mao}
\affiliation{Department of Astronomy, Tsinghua University, Beijing 100084, China}

\author[0000-0002-5854-8269]{Benjamin D. Wandelt}
\affiliation{Sorbonne Universit\'e, CNRS, UMR 7095, Institut d'Astrophysique de Paris (IAP), 98 bis bd Arago, 75014 Paris, France}
\affiliation{Sorbonne Universit\'e, Institut Lagrange de Paris (ILP), 98 bis bd Arago, 75014 Paris, France}
\affiliation{Center for Computational Astrophysics, Flatiron Institute, 162 5th Avenue, New York, NY 10010, USA}



\begin{abstract}

The first measurements of the 21~cm brightness temperature power spectrum from the epoch of reionization will very likely be achieved in the near future by radio interferometric array experiments such as the Hydrogen Epoch of Reionization Array (HERA) and the Square Kilometre Array (SKA). Standard MCMC analyses use an explicit likelihood approximation to infer the reionization parameters from the 21~cm power spectrum. In this paper, we present a new Bayesian inference of the reionization parameters where the likelihood is implicitly defined through forward simulations using density estimation likelihood-free inference (DELFI). Realistic effects including thermal noise and foreground avoidance are also applied to the mock observations from the HERA and SKA. We demonstrate that this method recovers accurate posterior distributions for the reionization parameters, and outperforms the standard MCMC analysis in terms of the location and size of credible parameter regions. With the minutes-level processing time once the network is trained, this technique is a promising approach for the scientific interpretation of future 21~cm power spectrum observation data. Our code {\tt 21cmDELFI-PS} is publicly available at \href{https://github.com/Xiaosheng-Zhao/21cmDELFI}{this link}. 

\end{abstract}



\section{Introduction} 
\label{sec:intro}

The cosmic 21~cm background from the epoch of reionization (EoR; \citealt{Furlanetto2006}) can provide direct constraints on the astrophysical processes regarding how H~{\small I} gas in the intergalactic medium (IGM) was heated and reionized by the first luminous objects (see, e.g. \citealt{2018PhR...780....1D}; \citealt{2022MNRAS.511.4005K}) that host ionizing sources. Observations of the 21~cm signal with radio interferometric array experiments, including the Precision Array for Probing the Epoch of Reionization (PAPER; \citealp{Parsons2010}), the Murchison Wide field Array (MWA; \citealp{Tingay2013}), the LOw Frequency Array (LOFAR; \citealp{Haarlem2013}), and the Giant Metrewave Radio Telescope (GMRT; \citealp{2017A&A...598A..78I}), have focused on the measurements of the power spectrum of the 21~cm brightness temperature fluctuations (hereafter ``21~cm power spectrum'') with stringent upper limits on it \citep{2013MNRAS.433..639P,2015ApJ...809...62P,2020MNRAS.493.1662M,2020MNRAS.493.4711T,2021arXiv210802263T}. In the near future, the first measurements of the 21~cm power spectrum from the EoR will very likely be achieved by the Hydrogen Epoch of Reionization Array (HERA; \citealp{DeBoer2017}) and the Square Kilometre Array (SKA; \citealp{Mellema2013}) with high signal-to-noise ratio. 

The 21~cm power spectrum is a two-point statistic that is sensitive to the parameters in the reionization models (hereafter ``reionization parameters''). To shed light on the astrophysical processes during reionization, posterior inference of reionization parameters from future 21~cm power spectrum measurements can be performed with the Monte Carlo Markov Chain (MCMC) sampling. In the standard MCMC analysis,  a multivariate Gaussian likelihood approximation is explicitly assumed, as in the publicly available code {\tt 21CMMC} \citep{2015MNRAS.449.4246G,2017MNRAS.472.2651G,Greig2018}\footnote{https://github.com/BradGreig/21CMMC}. Nevertheless, the predefined likelihood approximation may be biased, thereby misestimating the posterior distributions. 

To solve this problem,  simulation-based inference (SBI; \citealp{2019arXiv191013233P,Cranmer30055}), or  so-called ``likelihood-free inference'' (LFI), is proposed where the likelihood is implicitly defined through forward simulations. This allows building a sophisticated data model without relying on approximate likelihood assumptions. In the Approximate Bayesian Computation (ABC; \citealp{schafer2012likelihood,cameron2012approximate,hahn2017approximate}), the posterior distribution is approximated by adequate sampling of those parameters that are accepted if the ``distance'' between the sampling and the observation data meets some criterion. However, the convergence speed of the ABC method is slow in order to get the high-fidelity posterior distribution. 

Recently, machine learning has been extensively applied to 21~cm cosmology (e.g.\ \citealp{Shimabukuro2017, Kern2017Emulating,Schmit2018,Jennings2019,Gillet2019,hassan2019constraining,2021arXiv211203443Z, 2021arXiv211213866C,2022MNRAS.509.3852P,2022arXiv220108205S} and references therein). Specifically, \citet{Shimabukuro2017} and \citet{2019MNRAS.490..371D} applied the neural networks to the estimation of reionization parameters from the 21~cm power spectrum. It is worthwhile noting that such machine learning applications to 21~cm cosmology are mostly point estimate analyses, i.e.\ without posterior inference for recovered parameters. In \cite{2021arXiv210503344Z} (hereafter referred to as \citetalias{2021arXiv210503344Z}), we introduced the density estimation likelihood-free inference (DELFI; \citealp{alsing2018massive,alsing2019fast} and references therein) to the 21~cm cosmology, with which the posterior inference of the reionization parameters was performed for the first time from the three-dimensional tomographic 21~cm light-cone images. 
As a variant of LFI, DELFI contains various neural density estimators (NDEs) to learn the likelihood as the conditional density distribution of the target data given the parameters, from a number of simulated parameter-data pairs. It has been demonstrated to outperform the ABC method in terms of the convergence speed to get the high-fidelity posterior distribution \citep{alsing2018massive}. 

DELFI is a flexible framework to give the posterior inference of model parameters from data summaries. While the 21~cm power spectra have physical meaning as a summary statistic, from the DELFI point of view, these power spectra are just data summaries of the forward simulations. As such, in this paper, we will apply DELFI  in an amortized manner to the problem of posterior inference of reionization parameters from the 21~cm power spectrum. To mock the observations with the HERA and SKA, we will also take into account realistic effects including thermal noise and foreground avoidance. We will compare the results of DELFI with the standard MCMC analysis using {\tt 21CMMC}. To avoid the overconfidence in the SBI \citep{2021arXiv211006581H}, we will post-validate both marginal and joint posteriors from the SBI (\citealp{gneiting2007probabilistic,harrison2015validation,mucesh2021machine,2021arXiv210210473Z}) with statistical tests. To facilitate its application to future observation data, our code, dubbed {\tt 21cmDELFI-PS}, is made publicly available\footnote{https://github.com/Xiaosheng-Zhao/21cmDELFI}. 

The rest of this paper is organized as follows. We summarize DELFI in Section~\ref{sec:delfi}, and describe the data preparation in Section~\ref{sec:simu}, including the simulation of the 21~cm signal and the application of realistic effects. We present the posterior inference results and their validations in Section~\ref{sec:results}, and make concluding remarks in Section~\ref{sec:conclusion}. We leave the mathematical definitions of validation statistics to Appendix~\ref{sec: relia}, and the effect of the sample size to Appendix~\ref{sec: size}. 

\section{DELFI methodology}
\label{sec:delfi}

\begin{figure*}
	\includegraphics[width=\textwidth]{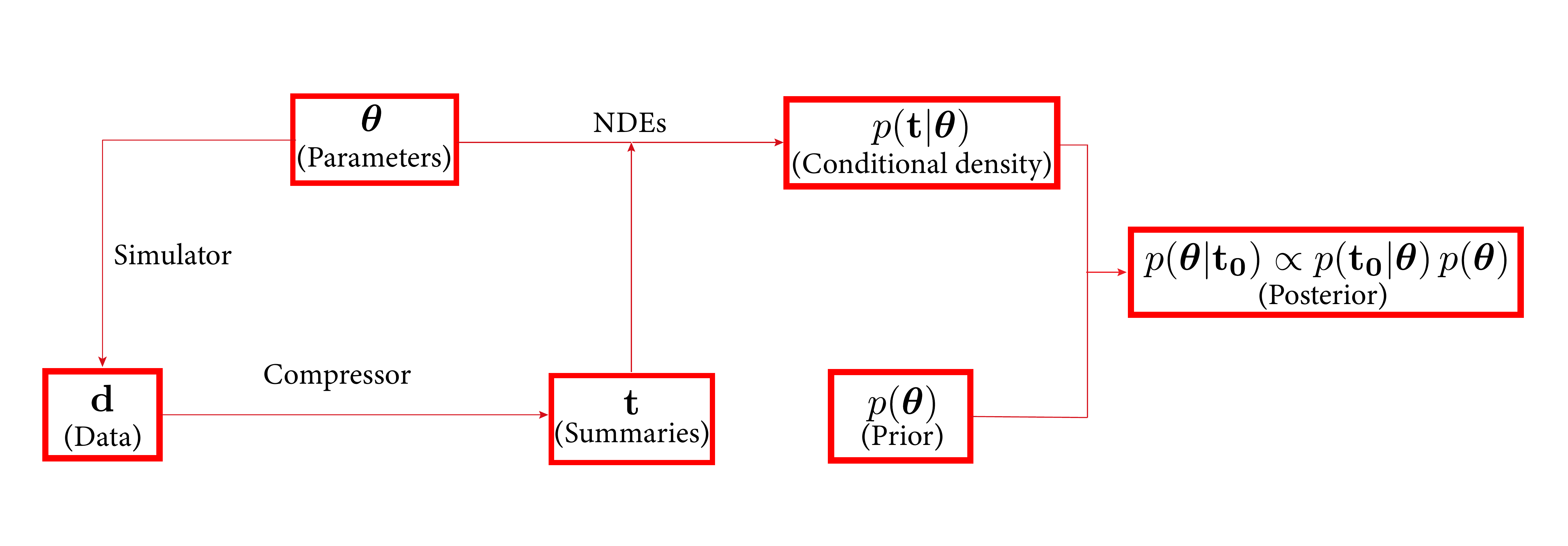}
    \caption{The workflow of {\tt 21cmDELFI-PS}. Here the ``simulator'' refers to {\tt 21cmFAST} that generates the 21~cm light-cone data cube (``data'') with the reionization parameters (``parameters''). The ``compressor'' refers to the the procedure of generating the 21~cm power spectra (``summeries'') from the data. The NDEs take the parameter-summary pairs ($\boldsymbol{\theta}, \mathbf{t}$) as the input and are trained to learn the conditional density $p(\mathbf{t} | \boldsymbol{\theta})$. The posterior distribution is inferred from the data likelihood evaluated at observation $\mathbf{t_{0}}$ and parameter prior, using  Bayes' Theorem. Figure adapted from Fig.~1 of \citetalias{2021arXiv210503344Z}.}
    \label{fig:pro}
\end{figure*}

We summarize DELFI in this section, and refer interested readers to \cite{alsing2018massive,alsing2019fast}; \citetalias{2021arXiv210503344Z} for details. DELFI is based on forward simulations that generate data $\mathbf{d}$ given parameters $\boldsymbol{\theta}$. If data vectors are of large dimensions, it is necessary to compress the data $\mathbf{d}$ into data summaries $\mathbf{t}$ that are of small dimension. DELFI contains various NDEs to learn the conditional density $p(\mathbf{t} | \boldsymbol{\theta})$ from a large number of simulated data pairs $\{\boldsymbol{\theta},\mathbf{t}\}$. NDEs that have been demonstrated to work include mixture density networks (MDN; \citealp{bishop1994mixture}) and masked autoregressive flows (MAF; \citealp{papamakarios2017masked}). With the conditional density, the likelihood $p(\mathbf{t_0} | \boldsymbol{\theta})$ can be evaluated at any data summary $\mathbf{t_0}$ from observed data. Then the posterior can be inferred using Bayes' Theorem, $p(\boldsymbol{\theta} | \mathbf{t_0}) \propto p(\mathbf{t_0} | \boldsymbol{\theta}) \, p(\boldsymbol{\theta})$, where $p(\boldsymbol{\theta})$ is the prior. The workflow of {\tt 21cmDELFI-PS} is illustrated in Fig.~\ref{fig:pro}.  

There are two choices in the application of DELFI --- amortized inference, and active-learning inference (aka multi-round inference). While it usually needs a large number of preprepared simulations for training the NDEs, amortized inference trains the {\it global} NDEs only once, and the inference from an observation is quick, so post-validations with many mock observations cost only a reasonable amount of computing time. On the other hand, active-learning inference focuses on the most probable region during inference and only trains the NDEs optimized in this local region, so it can effectively save the cost for simulations for inference with {\it one} observation data. However, this ``training during inference'' process should be repeated for inference with a new set of observational data, so it is computationally expensive to implement post-validations with many mock observations, each using active-learning inference. 
While the active-learning inference is a valid option in our code package {\tt 21cmDELFI-PS}, we choose to use the amortized inference in this paper, for the purpose of posterior post-validations. 

Our code {\tt 21cmDELFI-PS} employs the {\tt PYDELFI}\footnote{https://github.com/justinalsing/pydelfi} package for the DELFI implementation. We choose the MAFs (see \citetalias{2021arXiv210503344Z} for a detailed description) as the NDEs with fixed architectures that we find are flexible enough to handle all levels of datasets in {\tt 21cmDELFI-PS}. For all MAF architectures, we set two neural layers of a single transform, 50 neurons per layer, and five transformations in the MAFs. Ensembles of the MAFs are employed \citep{alsing2019fast,2021arXiv211006581H}, and the output posterior is obtained by stacking the posteriors from individual MAFs with weights according to their training errors. 

With the trained NDEs, it takes only about 5 minutes with a single core of an Intel Xeon Gold 6248 CPU (base clock speed 2.50 GHz) to process a mock observation and generate the posterior distribution. We plot the posterior contours with the {\tt emcee} module, and run 100 walkers for 1600 steps, with the first 600 steps dropped as ``burn-in''.

\section{Data Preparation}
\label{sec:simu}

\subsection{Cosmic 21~cm Signal}
The 21~cm brightness temperature at position ${\bf x}$ relative to the CMB temperature can be written \citep{Furlanetto2006} as 
\begin{equation}
T_{21}(\textbf{x},z)=\tilde{T}_{21}(z)\,x_{\rm HI}(\textbf{x})\,\left[1+\delta(\textbf{x})\right]\,(1-\frac{T_{\rm CMB}}{T_S})\,,\label{eqn:21cm}
\end{equation}
where $\tilde{T}_{21}(z) = 27\sqrt{[(1+z)/10](0.15/\Omega_{\rm m} h^2)}(\Omega_{\rm b} h^2/0.023)$ in units of mK. 
Here, $x_{\rm HI}({\bf x})$ is the neutral fraction, and $\delta({\bf x})$ is the matter overdensity, at position ${\bf x}$. We assume the baryon perturbation traces the cold
dark matter on large scales, so $\delta_{\rho_{\rm H}} = \delta$. 
In this paper, we focus on the limit where spin temperature $T_S \gg T_{\rm CMB}$, likely valid soon after reionization, though this assumption is strongly model dependent. As such, we can neglect the dependence on spin temperature. Also, as a demonstration of concept, we ignore the effect of peculiar velocity; such an effect can be readily incorporated in forward simulations by the algorithm introduced by \cite{2012MNRAS.422..926M}. 

In this paper, we use the publicly available code {\tt 21cmFAST}\footnote{https://github.com/andreimesinger/21cmFAST} \citep{Mesinger2007,Mesinger2011}, which can be used to perform semi-numerical simulations of reionization, as the simulator to generate the datasets. Our simulations were performed on a cubic box of 100 comoving Mpc on each side, with $66^3$ grid cells. Following the interpolation approach in \citetalias{2021arXiv210503344Z}, the snapshots at nine different redshifts of the same simulation box (i.e.\ with the same initial condition) are interpolated to construct a light-cone 21~cm data cube within the comoving distance of a simulation box along the line of sight (LOS). We concatenate 10 such light-cone boxes, each simulated with different initial conditions in density fields but with the same reionization parameters, together to form a full light-cone datacube of the size $100\times 100 \times 1000$ comoving ${\rm Mpc}^3$ (or $66\times 66 \times 660$ grid cells) in the redshift range $7.51 \le z \le 11.67$. To mimic the observations from radio interferometers, we subtract from the light-cone field the mean of the 2D slice for each 2D slice perpendicular to the LOS, because radio interferometers cannot measure the mode with ${\bf k}_\perp = 0$.

We divide the full light-cone 21~cm datacube into 10 light-cone boxes, each with the size of $(100\,{\rm cMpc})^3$  (or $66^3$ grid cells), and calculate the {\it light-cone} 21~cm power spectrum, defined by $\langle \widetilde{T_{21}}(\mathbf{k},z)\,\widetilde{T_{21}}(\mathbf{k^{\prime}},z)\rangle = (2\pi)^3 \delta(\mathbf{k}+\mathbf{k^{\prime}}) P_{21}(k,z)$. We also use the dimensionless 21~cm power spectrum, $\Delta_{21}^2(k,z) \equiv k^{3}P_{21}(k,z)/2\pi^{2}$. For each box, we choose to group the modes in Fourier space into 13 $k$-bins --- the upper bound of each $k$-bin is 1.35 times that of the previous bin. We then combine 10 such power spectra at different central redshifts into a single vector with the size of 130. 

We parametrize our reionization model as follows, and refer interested readers to \citetalias{2021arXiv210503344Z} for a detailed explanation of their physical meanings.

(1) $\zeta$, the ionizing efficiency, which is a combination of several parameters related to ionizing photons. In our paper, we vary $\zeta$ as $10 \le \zeta \le 250 $.

(2) $T_ { \mathrm { vir } }$, the minimum virial temperature of halos that host ionizing sources. In our paper, we vary this parameter as $ 4 \le \log _ { 10 } \left( T_{ \mathrm { vir } } / \mathrm { K } \right) \le 6 $.

Cosmological parameters are fixed in this paper as $\mathrm{( \Omega _ { \Lambda } , \Omega _ { m } , \Omega _ { b } , n_s , \sigma _ { 8 }} , h )=( 0.692,0.308,0.0484,0.968,0.815,$ $0.678 )$ \citep{ade2016planck}.

\begin{figure*}
	\centering
	\includegraphics[width=\textwidth]{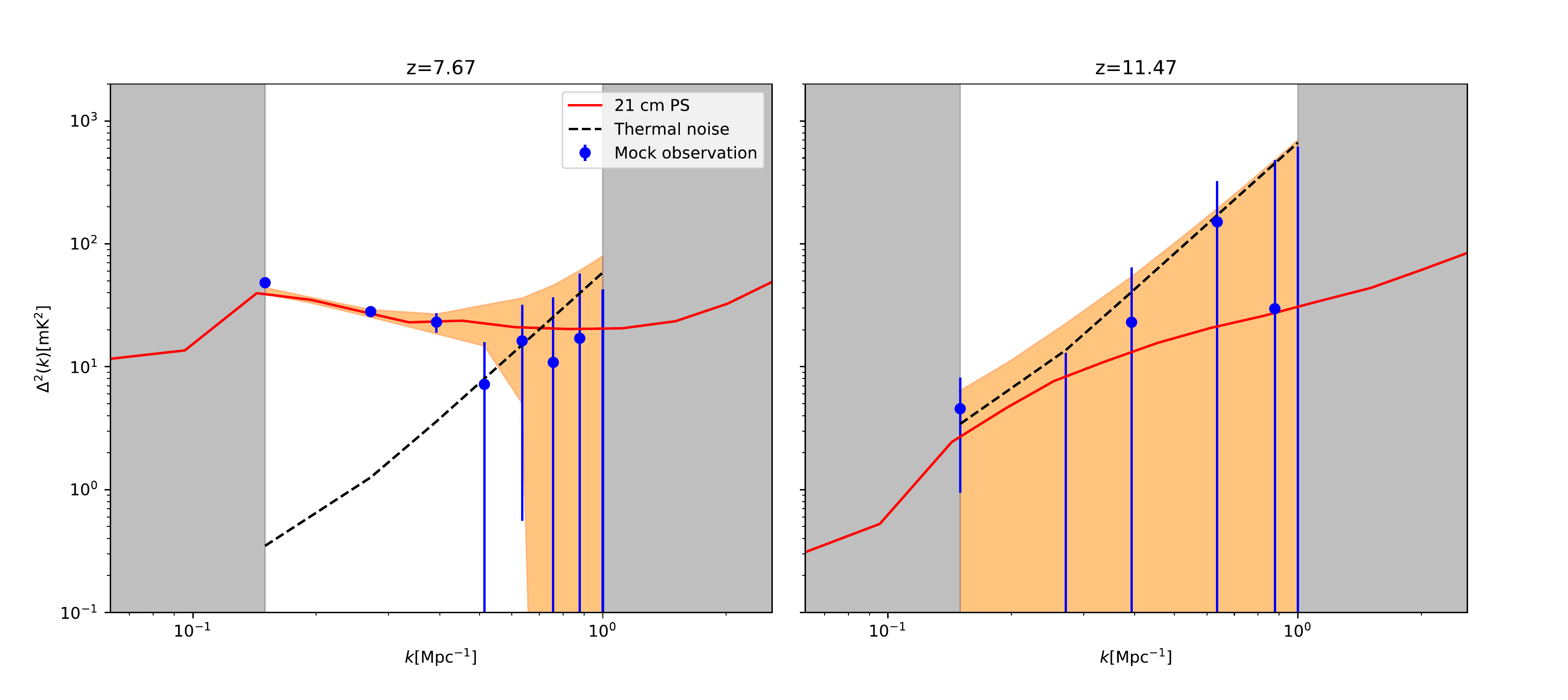}
    \caption{An example of the 21~cm power spectrum in the wavenumber range $0.15 \le k \le 1.0\, {\rm Mpc}^{-1}$ at $z=7.67$ (left) and $z=11.47$ (right). Shown are the cosmological light-cone 21~cm power spectrum from the {\tt 21cmFAST} simulation (red lines) with the reionization parameters defined in Table~\ref{tab: stat_a}, with the shaded orange regions around it representing the total noise power spectrum (including the contributions from thermal noise and sample variance errors), assuming the measurements with HERA. We also show the thermal noise (black lines) which dominates over the sample variance error, and the mock observed power spectrum (blue dots) with error bars representing the total noise. }
    \label{fig:sample}
\end{figure*}

\subsection{Thermal Noise}
\label{sec:instrument}

For the thermal noise estimation in this paper, we follow the treatment of the {\tt 21CMMC} code, for the purpose of comparison on the same ground. The {\tt 21CMMC} code employs the {\tt 21cmsense} module\footnote{https://github.com/steven-murray/21cmSense} \citep{Pober2013,Pober2014} to simulate the expected thermal noise power spectrum. We summarize the main assumptions here, and refer interested readers to  \cite{Pober2013,Pober2014,Greig2015,Greig2017} for details. 

The thermal noise power spectrum of any one mode of $uv$ pixels can be estimated as 
\begin{equation}
\Delta_{\mathrm{N}}^{2}(k) \approx X^{2} Y \frac{k^{3}}{2 \pi^{2}} \frac{\Omega^{\prime}}{2 t} T_{\mathrm{sys}}^{2},
\label{eqn:thermal_one_mode}
\end{equation}
where $X^{2} Y$ is a conversion factor converting observed bandwidths and solid angles to comoving volume in units of $(h^{-1}\ \mathrm{Mpc})^3$, $\Omega^{\prime}$ is a beam dependent factor \citep{2014ApJ...788..106P,Pober2014}, $t$ is the total integration time of all baselines on that particular $k$-mode. The system temperature $T_{\text {sys}} = T_{\text {rec}} + T_{\text {sky }}$, where $T_{\text {rec}}$ is the receiver temperature, and $T_{\text {sky}}$ is the sky temperature which can be modeled \citep{thompson2001synthesis} as $T_{\text {sky }}=60\left(\nu/300\,\mathrm{MHz}\right)^{-2.55} \mathrm{~K}$. 

The total noise power spectrum for a given $k$-mode combines the sample variance of the 21 cm power spectrum and the thermal noise using an inverse-weighted summation over all the individual measured modes \citep{Pober2013,Greig2017}, 
\begin{equation}
\delta \Delta_{\mathrm{T}+\mathrm{S}}^{2}(k)=\left(\sum_{i} \frac{1}{\left(\Delta_{\mathrm{N}, i}^{2}(k)+\Delta_{21}^{2}(k)\right)^{2}}\right)^{-1 / 2}\,,
\label{eqn:total_noise}
\end{equation}
where $\delta \Delta_{\mathrm{T}+\mathrm{S}}^{2}(k)$ is the total uncertainty from thermal noise and sample variance in a given $k$-mode, $\Delta_{\mathrm{N}, i}^{2}(k)$ is the per-mode thermal noise calculated with Equation (\ref{eqn:thermal_one_mode}) at each independent $k$-mode measured by the array as labelled by the index $i$, and $\Delta_{21}^{2}(k)$ is the cosmological 21~cm power spectrum (which contributes as the sample variance error here). 

In this paper for {\tt 21cmDELFI-PS}, we draw a random noise which follows the normal distribution $N(0, (\delta \Delta_{\mathrm{T}+\mathrm{S}}^{2}(k))^2)$, i.e.\ with zero mean and the variance of $(\delta \Delta_{\mathrm{T}+\mathrm{S}}^{2}(k))^2$, at each wavenumber $k$ for each sample, and add this noise to the cosmological 21~cm power spectrum in order to generate the mock 21~cm observed power spectrum. As an example, we illustrate the cosmological 21~cm power spectrum, thermal noise, and the mock observation in Fig.~\ref{fig:sample}. 

\begin{table}
	\centering
	\caption{Specifications for radio interferometric arrays.}
	\begin{tabular}{ccc} 
		\hline
		Parameter    & HERA & SKA \\
		\hline
        Telescope antennas                                                   & 331 & 224 \\ 
        Dish diameter ($\mathrm{m}$)                                & 14 & 35 \\ 
		Collecting area ($\mathrm{m^2}$)                                        & 50953 & 215513 \\
		$T_{\text {rec}} \mathrm{(K)}$                                                    & 100 & 0.1$T_{\text {sky}}+40$\\
		Bandwidth ($\mathrm{MHz}$)                                       & 8 & 8 \\
		Integration time ($\mathrm{h}$)                                                           & 1080 & 1080 \\
		\hline
	\end{tabular}
	\label{tab:instru parameters}
\end{table}

In this paper, we consider the mock observations with HERA and SKA, and follow \citet{Greig2015,Greig2017} for array specifications except for the minor changes in the SKA design as specified below. For both telescopes, we assume the drift-scanning mode with a total of 1080~h observation time \citep{Greig2015}. We list the key telescope parameters to model the thermal noise in Table~\ref{tab:instru parameters}. 

(1) HERA: we use the core design with 331 dishes forming a hexagonal configuration, with the diameter of 14~m for each dish \citep{beardsley2015adding}. 

(2) SKA\footnote{See the SKA1-low configurations in the latest technical document at \href{https://astronomers.skatelescope.org/wp-content/uploads/2016/09/SKA-TEL-SKO-0000422_02_SKA1_LowConfigurationCoordinates-1.pdf}{this link}.}: we only focus on the core area with 224 antenna stations randomly distributed within a core radius of about 500~m. The sensitivity improvements due to the arms of the SKA array distribution is negligible. 

\begin{table*}
	\centering
	\caption{Bayesian inference with {\tt 21cmDELFI-PS} and {\tt 21CMMC} for the ``Faint Galaxies Model''. Here, ``Pure signal'' refers to the mock observations of cosmological 21~cm power spectrum (i.e.\ without thermal noise or foreground avoidance); ``HERA'' (``SKA'') refers to the mock observations of the 21~cm power spectrum from HERA (SKA), which include the total noise (with the contributions from thermal noise and sample variance errors) and foreground avoidance. Note that the fractional error of an observable $A$ is related to the error of its common logarithm by $\Delta A/A \approx 2.3 \,\Delta\log_{10}(A)$. }
	\begin{tabular}{cccccccc}
	\hline\hline
			 & &\multicolumn{2}{c}{Pure signal} &\multicolumn{2}{c}{HERA} &\multicolumn{2}{c}{SKA} \\
			 \cmidrule(l{.75em}r{.75em}){3-4}
			 \cmidrule(l{.75em}r{.75em}){5-6}
			 \cmidrule(l{.75em}r{.75em}){7-8}
	Parameter & True value & {\tt 21cmDELFI-PS} & {\tt 21CMMC} & {\tt 21cmDELFI-PS} &  {\tt 21CMMC} & {\tt 21cmDELFI-PS} & {\tt 21CMMC}\\
		 \hline
		 $\log _ { 10 } \left( T_ { \rm vir }/{\rm K}\right)$ & 4.699 & $4.699^{+0.011}_{-0.011}  $& $4.673^{+0.033}_{-0.025}$ & $4.791^{+0.136}_{-0.114}  $ & $4.779^{+0.128}_{-0.109}$& $4.753^{+0.103}_{-0.095}$ & $4.758^{+0.117}_{-0.105}$\\
		 \hline
   $\log _ { 10 }(\zeta)$ & 1.477 & $1.481^{+0.011}_{-0.011}$ & $1.444^{+0.026}_{-0.023}$ & $1.560^{+0.098}_{-0.076}  $ & $1.533^{+0.099}_{-0.074}$ & $1.540^{+0.074}_{-0.065}$& $1.521^{+0.086}_{-0.073}$\\

	\hline
	\end{tabular}
	\label{tab: stat_a}
\end{table*}

\begin{table*}
	\centering
	\caption{Same as Table~\ref{tab: stat_a} but for the ``Bright Galaxies Model''.}
	\begin{tabular}{cccccccc}
	\hline\hline
			 & &\multicolumn{2}{c}{Pure signal} &\multicolumn{2}{c}{HERA} &\multicolumn{2}{c}{SKA} \\
			 \cmidrule(l{.75em}r{.75em}){3-4}
			 \cmidrule(l{.75em}r{.75em}){5-6}
			 \cmidrule(l{.75em}r{.75em}){7-8}
	Parameter & True value & {\tt 21cmDELFI-PS} & {\tt 21CMMC} & {\tt 21cmDELFI-PS} &  {\tt 21CMMC} & {\tt 21cmDELFI-PS} & {\tt 21CMMC}\\
		 \hline
		 $\log _ { 10 } \left( T_ { \rm vir }/{\rm K}\right)$ & 5.477 & $5.480^{+0.015}_{-0.016} $& $5.435^{+0.050}_{-0.052}$ & $5.364^{+0.097}_{-0.117} $ & $5.375^{+0.104}_{-0.144}$& $5.391^{+0.089}_{-0.113}$ & $5.379^{+0.102}_{-0.131}$\\
		 \hline
   $\log _ { 10 }(\zeta)$ & 2.301 & $2.306^{+0.023}_{-0.023}$ & $2.226^{+0.077}_{-0.072}$ & $2.186^{+0.126}_{-0.145} $ & $2.159^{+0.133}_{-0.168}$ & $2.211^{+0.115}_{-0.141}$& $2.161^{+0.136}_{-0.159}$\\

	\hline
	\end{tabular}
	\label{tab: stat_b}
\end{table*}

\subsection{Foreground Cut}
\label{subsec:foreground}

To remove the bright radio foreground, we adopt the ``moderate''  foreground avoidance strategy in the {\tt 21cmSense} module. This strategy \citep{Pober2014} avoids the foreground ``wedge'' in the cylindrical $(k_{\perp},\ k_{\parallel})$ space, where the ``wedge'' is defined to extend $0.1 \ h\ \mathrm{Mpc^{-1}}$ beyond the horizon limit (with the slope of about 3.45 at $z=8$). This also incorporates the coherent addition of all baselines for a given $k$-mode.

\subsection{Database}

We use the Latin Hypercube Sampling to scan the reionization parameter space. For the mock observations with only cosmological 21~cm power spectrum, we generate 18000 samples with different reionization parameters used for training the NDEs, and 300 additional samples for testing or validating the DELFI. For the mock observations with noise and foreground cut, we generate 9000 samples with different reionization parameters, and make 10 realizations of total noise and foreground cut for each such sample --- because the noise is random --- so totally 90000 samples are used for training the NDEs, and 300 additional samples are for testing the DELFI. 
We discuss the effect of the sample size on the inference performance in Appendix~\ref{sec: size}. The initial conditions for all samples (with different reionization parameters) were independently generated by sampling spatially correlated Gaussian random fields with the matter power spectrum given by linear theory.

\subsection{The {\tt 21CMMC} Setup}

For the purpose of comparison, we also run the {\tt 21CMMC} code. We summarize the main setup of {\tt 21CMMC} here, and refer interested readers to \citetalias{2021arXiv210503344Z} for the detail. We generate the mock power spectra at 10 different redshifts, each estimated from a coeval box of 100 comoving Mpc on each side. Strictly speaking, power spectra from the {\it light-cone} boxes should be employed in both mock observation and MCMC sampling for {\tt 21CMMC}, because the {\tt 21cmDELFI-PS} is based on the light-cone datacube. However, the {\tt 21CMMC} analysis from the coeval boxes does not significantly change the inference results, since the light-cone effect is only non-negligible at large scales. Since the {\tt 21CMMC} analysis is not the focus of our paper but just serves for comparison, we choose to use coeval boxes in both mock observation and MCMC sampling for {\tt 21CMMC} herein, for self-consistency, to save computational time. This at least avoids the bias caused by using the light-cone box for the mock observation and the coeval box in the MCMC sampling \citep{Greig2018}. 

In the case with the thermal noise and foreground cut, the total noise power spectrum in Equation~(\ref{eqn:total_noise}) is used as the variance of the likelihood function, to include the noise for {\tt 21CMMC}. In the hypothetical case where there is no thermal noise or foreground in the mock observations, the likelihood function only includes the sample variance from the mock observation, $P_{\rm sv} = P_{21}(k) / \sqrt{N(k)}$, where $N(k)$ is the number of modes in the spherical shell of $k$-bin. We turn off the modelling uncertainty parameter in {\tt 21CMMC} that parametrizes the systematics in the semi-numerical simulations. 

We perform the Bayesian inference with 200 walkers for the case of only cosmological 21~cm signal. For each walker, we choose the ``burn in chain'' number to be 250, and the main chain number to be 3000. (See the tests in \citetalias{2021arXiv210503344Z}.) For the case with the realistic effects, we employ only 100 walkers with same other settings, because of the convergence of the {\tt 21CMMC} results. 


\begin{figure*}
	\centering
	\includegraphics[width=\textwidth]{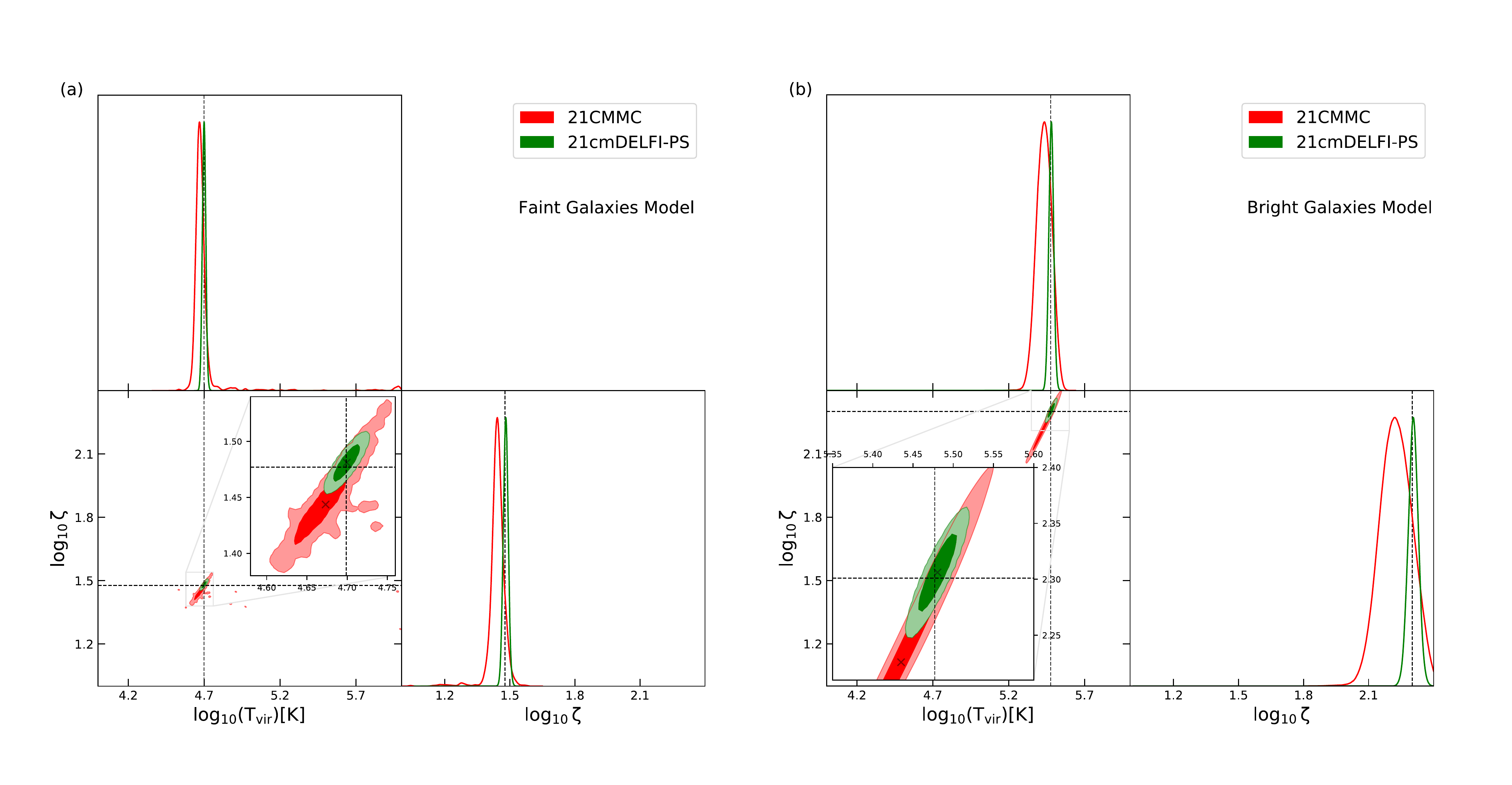}
    \caption{The posteriors estimated from the cosmological 21~cm power spectrum (i.e.\ without thermal noise or foreground avoidance) by two different approaches --- {\tt 21cmDELFI-PS} (green) and {\tt 21CMMC} (red), for two mock observations, the ``Faint Galaxies Model'' (left) and the ``Bright Galaxies Model'' (right). We show the median (cross), the $1\sigma$ (dark) and $2\sigma$ (light) confidence regions. The dashed lines indicate the true parameter values. }
    \label{fig:pure_2}
\end{figure*}

\begin{figure*}
	\centering
	\includegraphics[width=\textwidth]{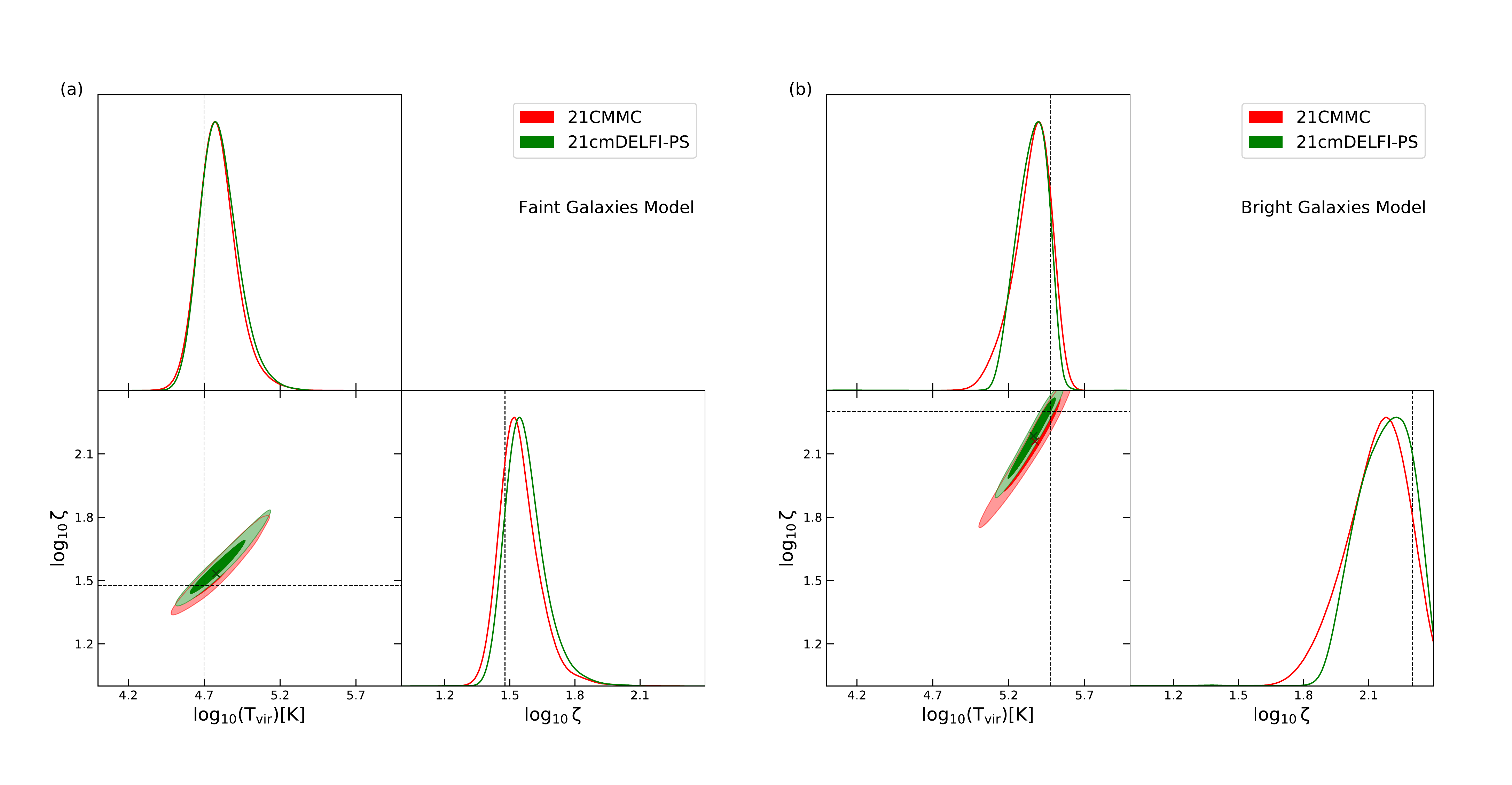}
    \caption{Same as Fig.~\ref{fig:pure_2} but the estimations are made from mock observations of the 21~cm power spectrum from HERA, which include the total noise (with the contributions from thermal noise and sample variance errors) and foreground avoidance. }
    \label{fig:noised_hera_2}
\end{figure*}

\begin{figure*}
	\centering
	\includegraphics[width=\textwidth]{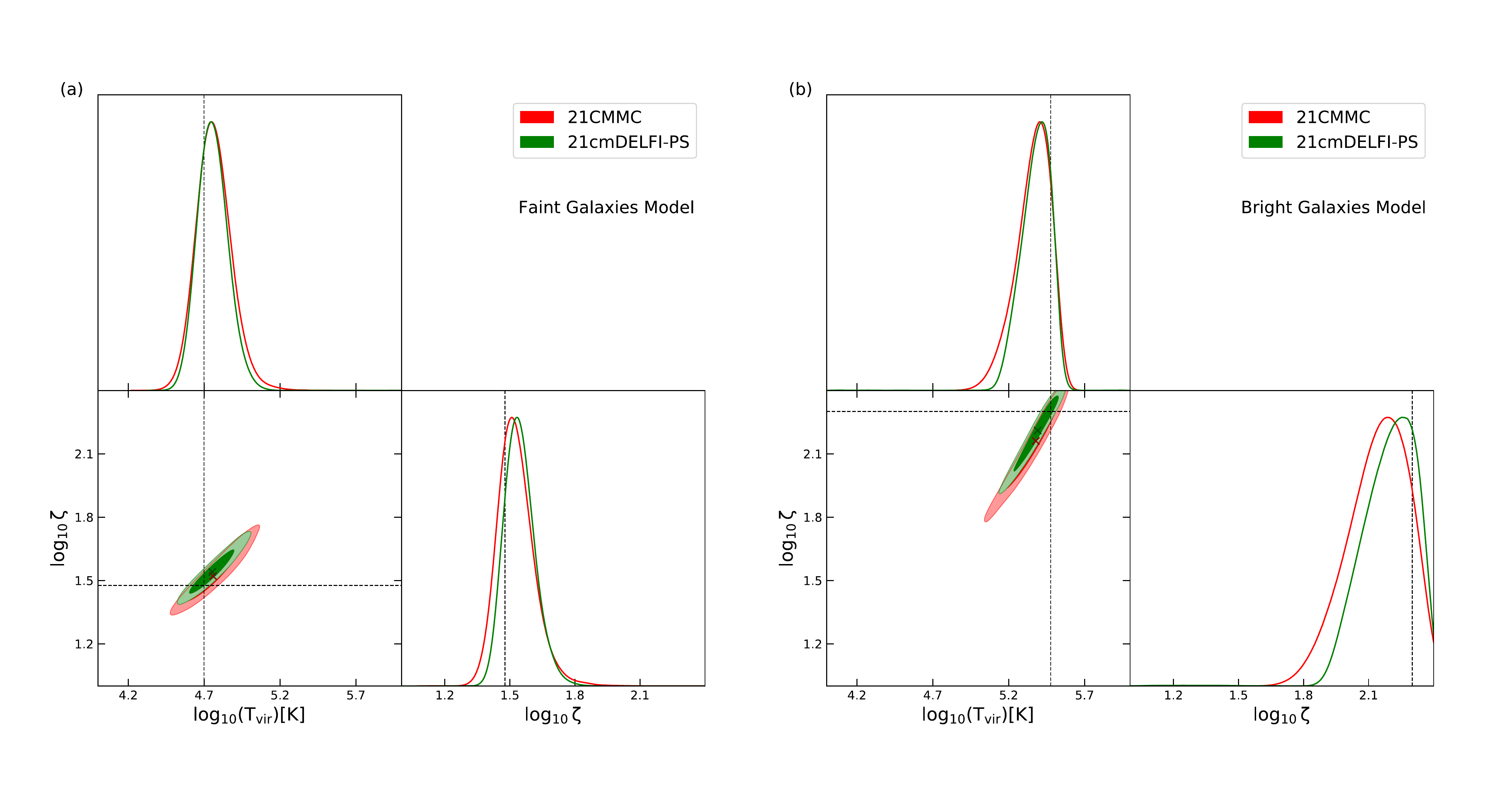}
    \caption{Same as Fig.~\ref{fig:noised_hera_2} but with mock observations of the 21~cm power spectrum from SKA.}
    \label{fig:noised_ska_2}
\end{figure*}

\begin{figure*}
\centering
	\includegraphics[width=0.9\textwidth]{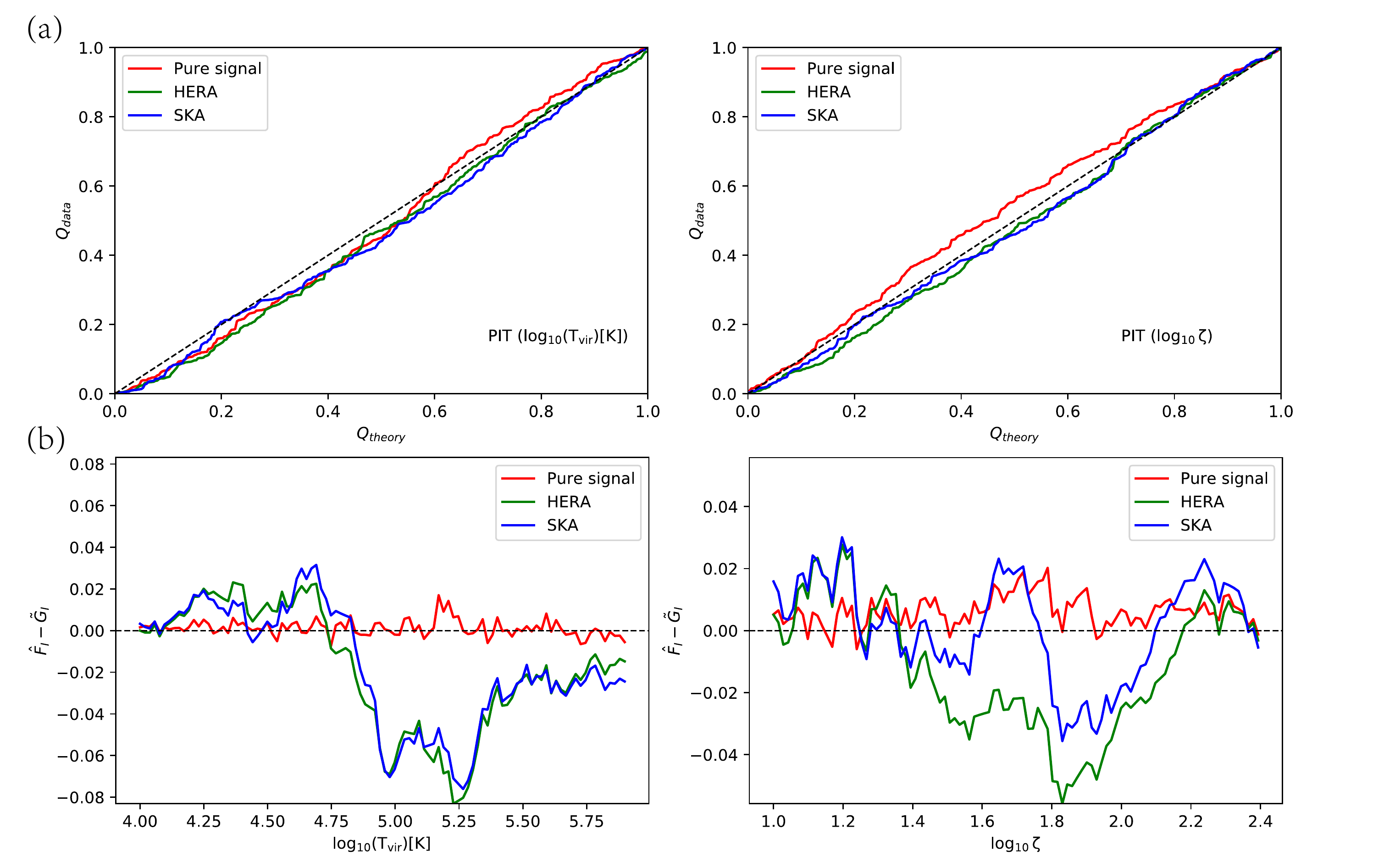}
    \caption{Validation of marginal posteriors. (a) Quantile-quantile plot, i.e.\ quantile of the PIT distribution from the test sample data vs that from a theoretical uniform PIT distribution. The diagonal dashed line represents the ideal case ($Q_{\rm data} = Q_{\rm theory}$). 
    (b) Marginal calibration, i.e.\ $\hat{F}_{I}(\theta) - \tilde{G}_{I}(\theta)$. The diagonal dashed line represents the ideal case (identically zero). 
    We show the results for $\theta = \log _ { 10 } \left( T_ { \rm vir } \right)$ (left panels) and $\log _ { 10 } \zeta$ (right panels), respectively, and for mock observations with only cosmological 21~cm signal (``Pure signal''; red), HERA (green) and SKA (blue), respectively. }
    \label{fig:PIT}
\end{figure*}

\section{Results}
\label{sec:results}

\subsection{Posterior Inference for Mock Observations}

In this section, we test the Bayesian inference by {\tt 21cmDELFI-PS} and compare its results with {\tt 21CMMC}. As a demonstration of concept, we consider two representative mock observations, the ``Faint Galaxies Model'' and ``Bright Galaxies Model'', whose definitions are listed as the ``True value'' in Table~\ref{tab: stat_a} and Table~\ref{tab: stat_b}, respectively, following \citet{2017MNRAS.472.2651G}. These models are chosen as two examples with extreme parameter values. Their global reionization histories are similar, but reionization in the ``Faint Galaxies Model'' is powered by more abundant low-mass galaxies yet with smaller ionization efficiency (due to smaller escape fraction of ionizing photons) than in the ``Bright Galaxies Model'', so the H~{\small II} bubbles in the former are smaller and more fractal than in the latter. In Table~\ref{tab: stat_a} and Table~\ref{tab: stat_b}, we list the median and $1\sigma$ errors of posterior inference for both mock observations, respectively. 

In the hypothetical case where there is no thermal noise or foreground in the mock observations of the cosmological 21~cm power spectrum, we show the results of posterior inference for both mock observations in Fig.~\ref{fig:pure_2}. Both {\tt 21cmDELFI-PS} and {\tt 21CMMC} can recover posterior distributions for the reionization parameters in the sense that the medians are within the estimated $1\sigma$ confidence region. However, {\tt 21cmDELFI-PS} outperforms {\tt 21CMMC} in terms of the location and size of credible parameter regions. Quantitatively, for the ``Faint Galaxies Model'', the systematic shift (i.e.\ relative errors of the predicted medians with respect to the true values) and the $1\sigma$ statistical errors are $0\%\pm 0.23\%$ ($-0.55\% {+0.70\% \atop -0.53\%}$) for $\log _ { 10 } \left( T_ { \rm vir } \right)$ with {\tt 21cmDELFI-PS} ({\tt 21CMMC}), respectively, and $0.27\%\pm 0.74\%$ ($-2.2\%{+1.8\% \atop -1.6\%}$) for $\mathrm{log_{10}\zeta}$ with {\tt 21cmDELFI-PS} ({\tt 21CMMC}), respectively. Not only are the predicted medians of {\tt 21cmDELFI-PS} much closer to the true values than {\tt 21CMMC}, but also the estimated statistical errors in the former are $2 - 3$ times smaller than in the latter. These results hold generically for the ``Bright Galaxies Model'', too. In {\tt 21CMMC}, an explicit likelihood assumption is made, namely that the likelihood is a multivariate Gaussian, with independent measurements at each redshift and at each $k$-mode. Our results question the validity of this explicit likelihood assumption in the hypothetical case where there is no thermal noise or foreground in the mock 21~cm data. Indeed, \cite{10.1093/mnras/stw2599,10.1093/mnras/stz1561,10.1093/mnras/staa2090} show that the non-Gaussianity in the covariance of the 21~cm power spectrum is non-negligible. 

Now we apply the realistic effects including the total noise power spectrum (with thermal noise and sample variance errors) as well as using foreground cut to avoid the foreground. In Figs.~\ref{fig:noised_hera_2} and \ref{fig:noised_ska_2}, we show the results of posterior inference for mock observations with HERA and SKA, respectively. We find that both {\tt 21cmDELFI-PS} and {\tt 21CMMC} can recover posterior distributions for the reionization parameters in this case. Also, the performances with HERA and SKA are comparable, which was also found in \citet{Greig2015,Greig2017}. For mock observations with both HERA and SKA, the statistical (fractional) errors are $\sim 2-3\%$ for $\log _ { 10 } \left( T_ { \rm vir } \right)$ and $\sim 5-7\%$ for $\mathrm{log_{10}\zeta}$, or equivalently $\sim 23-33\%$ for $T_ { \rm vir }$ and $\sim 17-39\%$ for $\zeta$. 

Comparing {\tt 21cmDELFI-PS} and {\tt 21CMMC}, their recovered medians are in good agreement, but the $1\sigma$ credible regions estimated by {\tt 21cmDELFI-PS} are in general slightly smaller than those by {\tt 21CMMC}. This agreement suggests that the explicit Gaussian likelihood assumption is approximately valid when thermal noise and foreground cut are incorporated in the mock 21~cm data. This is reasonable because thermal noise dominates over the cosmic variance in HERA and SKA, and thus the non-Gaussian covariance due to the cosmic variance is subdominant. Also, we find that the direction of degeneracies in the parameter space are almost the same for these two codes, which reflects the fact that both codes use the same simulator {\tt 21cmFAST}, so the dependency of the 21~cm power spectrum on the reionization parameters is the same. 

It is worthwhile noting that the mock observed power spectrum as the input of {\tt 21CMMC} is the cosmological 21~cm signal from the {\tt 21cmFAST} simulation (i.e.\ the red line in Fig.~\ref{fig:sample}), as the default setting of {\tt 21CMMC}. If we apply the noise to the mock observed power spectrum (i.e.\ the blue dots in Fig.~\ref{fig:sample}) as the input of {\tt 21CMMC}, the inference performance of {\tt 21CMMC} can be degraded significantly. However, this is technically solvable by generating multiple noise realizations in the MCMC sampling, albeit with considerably larger computational cost. In comparison, the input of {\tt 21cmDELFI-PS} is the mock with noise, in which case the Bayesian inference works well. This is because {\tt 21cmDELFI-PS} learns the effect of variations due to noise by including 10 noise realizations for each reionization model in the training samples, with reasonable computational time for training.


\begin{figure*}
\centering
    \includegraphics[width=\textwidth]{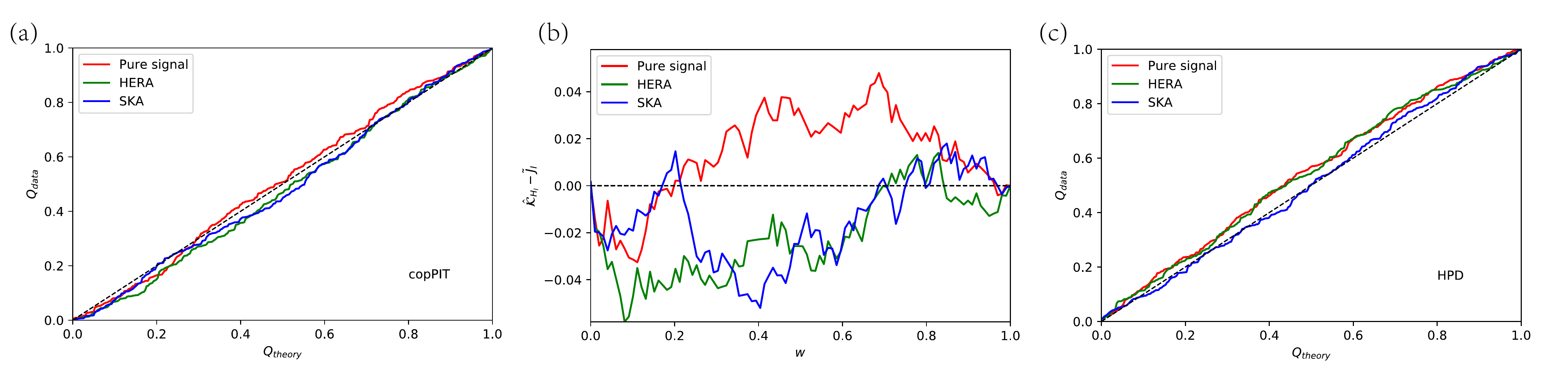}
    \caption{Joint posteriors validation. 
    (a) Quantile-quantile plot, i.e.\ quantile of the copPIT distribution from the test sample data vs that from a theoretical uniform copPIT distribution. 
    (b) Kendall calibration, i.e.\ $\hat{\mathcal{K}}_{H_{I}}(w) - \tilde{J}_{I}(w)$, where $w$ is a variable between zero and unity. The diagonal dashed line represents the ideal case (identically zero). 
    (c) Quantile-quantile plot but for the quantile of the HPD distribution. 
    We show the results for mock observations with only cosmological 21~cm signal (``Pure signal''; red), HERA (green) and SKA (blue), respectively. 
    In Panels (a) and (c), the diagonal dashed line represents the ideal case ($Q_{\rm data} = Q_{\rm theory}$). }
    \label{fig:calib_2d}
\end{figure*}

\begin{table}
	\centering
	\caption{The p-values for the null hypotheses that these statistics are of a uniform distribution.}
	\begin{tabular}{ccccccc}
	\hline\hline
			 &\multicolumn{2}{c}{Pure signal}&\multicolumn{2}{c}{HERA}&\multicolumn{2}{c}{SKA}\\
			 \cmidrule(l{.75em}l{.75em}r{.75em}){2-3}
			 \cmidrule(l{.75em}l{.75em}r{.75em}){4-5}
			 \cmidrule(l{.75em}l{.75em}r{.75em}){6-7}
	Statistics & KS & CvM & KS & CvM & KS & CvM \\
		 \hline
		 PIT ($\log _ { 10 } \left( T_ { \rm vir } \right)$)    & 0.37 & 0.13 & 0.16 & 0.09 & 0.11 & 0.10 \\
		 \hline
		 PIT ($\log _ { 10 } \zeta$) &0.14  & 0.05 & 0.30  &0.20 &0.36 &0.36\\
	     \hline
		 copPIT        &0.54&0.39& 0.13 & 0.17 &0.22 &0.29\\
		 \hline
		 HPD           &0.04 & 0.01  & 0.03 & 0.02 &0.73&0.67\\
	\hline
	\end{tabular}
	\label{tab: bayesian num}
\end{table}

\subsection{Validation of the posterior}

In this subsection, we perform the validation\footnote{In the convention of statistics, these tools are called ``calibration''. In our context of testing the posteriors, however, the term ``validation'' is more appropriate.} of posteriors, which tests statistically the accuracies of inferred posterior distributions. We employ 300 samples for mock observations with only cosmological 21~cm power spectrum and with realistic effects with HERA and with SKA, respectively. These samples are randomly chosen from the allowed region in the parameter space in which the neutral fraction satisfies $0.08 \le x_{\rm HI} \le 0.81$ at $z=7.1$, corresponding to the $2 \sigma$ confidence region from the IGM damping wing of ULASJ1120+0641 \citep{2017MNRAS.466.4239G}. The mathematical definitions of validation statistics are left to Appendix~\ref{sec: relia}. 

(1) Validation of marginal posteriors. We first focus on the validation of posteriors for each single parameter marginalized over other parameters. In Fig.~\ref{fig:PIT}, we perform two statistics --- (i) quantile of the probability integral transform (PIT), and (ii) $\hat{F}_{I}(\theta) - \tilde{G}_{I}(\theta)$, where $\hat{F}_{I}(\theta)$ is the predictive cumulative distribution function (CDF) and $\tilde{G}_{I}(\theta)$ is the empirical CDF, for $\theta = \log _ { 10 } \left( T_ { \rm vir } \right)$ and $\log _ { 10 } \zeta$. In the quantile-quantile (QQ) plot, we find that the curves for all cases of mock observations are close to the diagonal line ($Q_{\rm data} = Q_{\rm theory}$). In the marginal calibration, we find that the values of $\hat{F}_{I}(\theta) - \tilde{G}_{I}(\theta)$ for all cases of mock observations are small (less than $0.1$). These findings meet the expectations with which the marginal posterior distributions are probabilistically calibrated. 

(2) Joint posteriors validation. Next, we focus on the validation of posteriors in the joint parameter space. In Fig.~\ref{fig:calib_2d}, we consider three statistics --- (i) quantile of the copula probability integral transformation (copPIT); (ii) $\hat{\mathcal{K}}_{H_{I}}(w) - \tilde{J}_{I}(w)$, where $\hat{\mathcal{K}}_{H_{I}}(w)$ is the average Kendall distribution function and $\tilde{J}_{I}(w)$ is the empirical CDF of the joint CDFs, and here $w$ is a variable between zero and unity; (iii) the highest probability density (HPD). In both QQ plots for the copPIT and for the HPD, we find that the curves for all cases of mock observations are close to the diagonal line ($Q_{\rm data} = Q_{\rm theory}$). In the Kendall calibration, we find that the value of $\hat{\mathcal{K}}_{H_{I}}(w) - \tilde{J}_{I}(w)$ for all cases of mock observations are small (less than $0.1$). These findings meet the expectations with which the joint posterior distributions are probabilistically copula calibrated and probabilistically HPD calibrated.

(3) Hypothesis tests. The aforementioned validations of marginal and joint posteriors provide qualitative measures of the validation of the inferred posteriors. The quantitative metrics for the uniformity of distribution of these statistics (PIT, copPIT and HPD) are in the hypothesis tests. For both marginal posteriors and joint posteriors, we perform the Kolmogorov-Smirnov (KS) test and the Cram\'er-von Mises (CvM) test. The $p$-value of KS (CvM) test is the probability to obtain a sample data distribution with $D$ ($D^*$) larger than the measured $D_{\rm obs}$ ($D^*_{\rm obs}$). The difference of these two metrics is that the KS measure $D$ is sensitive to the median, but the CvM measure $D^*$ incorporates the information of the tails of a distribution. In Table~\ref{tab: bayesian num}, we list the $p$-values for the PIT of two individual reionization parameters (for the validation of marginal posteriors), and those for the copPIT and the HPD (for the joint posteriors validation). In all cases of mock observations, we find that the $p$-value for the PIT and for the copPIT is larger than 0.05, and that for the HPD is larger than 0.01. This means that at the significance level of $5\%$/$5\%$/$1\%$, the null hypothesis that the PIT/copPIT/HPD distribution is uniform is accepted. In other words, the posteriors are probabilistically calibrated, probabilistically copula calibrated, and probabilistically HPD calibrated. 

In sum, the posteriors obtained with {\tt 21cmDELFI-PS} are valid under all statistical tests of marginal and joint posteriors.

\section{Summary}
\label{sec:conclusion}

In this paper, we present a new Bayesian inference of the reionization parameters from the 21~cm power spectrum. Unlike the standard MCMC analysis that uses an explicit likelihood approximation, in our approach, the likelihood is implicitly defined through forward simulations using DELFI. In DELFI, once the neural density estimators are trained to learn the likelihood, the inference speed for a test observation of power spectrum is very fast --- about $5$ minutes with a single core of an Intel CPU (base clock speed 2.50 GHz).  

We show that this method (dubbed {\tt 21cmDELFI-PS}) recovers accurate posterior distributions for the reionization parameters, using mock observations without and with realistic effects of thermal noises and foreground cut with HERA and SKA. For the purpose of comparison, we perform MCMC analyses of the 21~cm power spectrum using a Gaussian likelihood approximation. We demonstrate that this new method ({\tt 21cmDELFI-PS}) outperforms the standard MCMC analysis (using {\tt 21CMMC}) in terms of the location and size of credible parameter regions, in all cases of mock observations. We also perform the validation of both marginal and joint posteriors with sophisticated statistical tests, and demonstrate that the posteriors obtained with {\tt 21cmDELFI-PS} are statistically self-consistent. 

It is interesting to note that in the scenario of only cosmological 21~cm signal without thermal noises and foreground contamination, the 21~cm power spectrum analysis with {\tt 21cmDELFI-PS} even outperforms our previous analysis from 3D tomographic 21~cm light-cone images with {\tt DELFI-3D CNN} \citepalias{2021arXiv210503344Z}, where 3D CNN was applied to compress the light-cone images into low-dimensional summaries. Since the power spectrum only contains a subset of information in images, this implies that 3D CNN is not the optimal imaging compressor. We leave it to a followup of \citetalias{2021arXiv210503344Z} to search for a new imaging compressor with better performance results than 3D CNN and {\tt 21cmDELFI-PS}.

The development of this new Bayesian inference method is timely because the first measurements of the 21~cm power spectrum from the EoR will very likely be achieved in the near future by HERA and SKA. The DELFI framework is flexible for incorporating more realistic effects through forward simulations, so this technique will be a promising approach for the scientific interpretation of future 21 cm power spectrum observation data. To facilitate its application, we have made {\tt 21cmDELFI-PS} publicly available. 

As a demonstration of concept, this paper only considers the limit where spin temperature $T_S \gg T_{\rm CMB}$ and therefore neglects the dependence on spin temperature. We leave it to future work to extend the parameter space to the cosmic dawn parameters that affect the IGM heating and Ly$\alpha$-pumping. Also, Bayesian inference of the reionization parameters from other summary statistics, e.g.\ 21~cm bispectrum \citep{2022MNRAS.510.3838W}, can be developed in a similar manner to {\tt 21cmDELFI-PS}, because these statistics are just data summaries from the DELFI point of view. We leave it to future work to make such developments. 

\section*{Acknowledgements}
This work is supported by National SKA Program of China (grant No.~2020SKA0110401), NSFC (grant No.~11821303), and National Key R\&D Program of China (grant No.~2018YFA0404502). 
BDW acknowledges support from the Simons Foundation.
We thank Paulo Montero-Camacho, Jianrong Tan, Steven Murray, Nicholas Kern and Biprateep Dey for useful discussions and helps, and the anonymous referee for constructive comments. We acknowledge the Tsinghua Astrophysics High-Performance Computing platform at Tsinghua University for providing computational and data storage resources that have contributed to the research results reported within this paper.

\software{21CMMC \citep{2015MNRAS.449.4246G,2017MNRAS.472.2651G,Greig2018}, 21cmFAST \citep{Mesinger2007,Mesinger2011}, pydelfi \citep{alsing2019fast}, TensorFlow \citep{abadi2016tensorflow}, GetDist \citep{Lewis:2019xzd}, NumPy \citep{harris2020array}, Matplotlib \citep{Hunter:2007}, SciPy \citep{2020SciPy-NMeth}, scikit-learn \citep{Scikit-learn}, Python2 \citep{van1995python}, Python3 \citep{py3}, galpro \citep{mucesh2021machine}, seaborn \citep{Waskom2021}, Astropy \citep{2013A&A...558A..33A, 2018AJ....156..123A}.}

\appendix

\section{Statistical Tools for Validations} 
\label{sec: relia}

In this section, we introduce some statistical tools for validation of the marginal or joint posterior distributions inferred from data, following \cite{gneiting2007probabilistic,10.1214/14-EJS964,harrison2015validation,mucesh2021machine}. 

\subsection{Validation of Marginal Posteriors}

The tools in this subsection focus on the validation of posteriors for each single parameter marginalized over other parameters. 

\subsubsection{Probabilistic Calibration}
Consider an inferred marginal distribution $f(\theta)$, where $\theta$ is a marginalized parameter. For example, $\theta = \log _ { 10 } \left( T_ { \rm vir } \right)$ or $\mathrm{log_{10}\zeta}$ in this paper, and $f(\theta)$ is the probability distribution function that is generated in the MCMC chain given the observed data. 
We define the probability integral transform (PIT) as the cumulative distribution function (CDF) of this marginal distribution \citep{gneiting2007probabilistic,mucesh2021machine}
\begin{equation}
	{\rm PIT}\,(\tilde{\theta}) \equiv \int_{-\infty}^{\tilde{\theta}} f(\theta) \,\mathrm{d} \theta\,,
\end{equation}
where $\tilde{\theta}$ is the true value given the data. The marginal distributions are probabilistically calibrated if true values are randomly drawn from the real distributions. This is equivalent to the statement that the distribution of PIT is uniform. In the so-called quantile-quantile (QQ) plot, wherein the quantile of the PIT distribution from the data is compared with that from a uniform distribution of PIT, then this QQ plot falls on the diagonal line if the PIT distribution is exactly uniform.  

\subsubsection{Marginal Calibration} 

The uniformity of PIT distribution is only a necessary condition for real marginal posteriors \citep{hamill2001interpretation,gneiting2007probabilistic,mucesh2021machine,2021arXiv210210473Z}. As a complementary test, 
the marginal calibration \citep{gneiting2007probabilistic,mucesh2021machine} compares the average predictive CDF and the empirical CDF of the parameter.  The average predictive CDF is defined as 
\begin{equation}
\hat{F}_{I}(\theta)\equiv\frac{1}{N} \sum_{i=1}^{N} F_{i}(\theta)\,,
\end{equation}
where $N$ is the number of test samples, and $F_i(\theta)$ is the CDF of the posterior distribution of the parameter given the data of the $i$th sample. 
The empirical CDF is defined as 
\begin{equation}
\tilde{G}_{I}(\theta)\equiv\frac{1}{N} \sum_{i=1}^{N} \mathbbm{1}\left\{\tilde{\theta}_i \leq \theta \right\}\,,
\end{equation}
where the indicator function $\mathbbm{1}\left\{ A \right\}$ returns to be unity if the condition $A$ is true, and zero otherwise. 

If the posteriors are marginally calibrated, $\hat{F}_{I}(\theta)$ and $\tilde{G}_{I}(\theta)$ should agree with each other for any parameter $\theta$, so their difference should be small. 

\subsubsection{Hypothesis Tests}

To provide a quantitative measure of the similarity between the distribution of PIT and a uniform distribution, we adopt two metrics, the Kolmogorov-Smirnov (KS; \citealp{Kolmogorov1992}) test and Cram\'er-von Mises (CvM; \citealp{10.1214/aoms/1177704477}) test. The classical KS test is based on the distance measure $D$ defined by
\begin{equation}
D \equiv \max _{x}\left|F_{N}(x)-F(x)\right|,
\label{eqn:KS}
\end{equation}
where $F_{N}(x)$ is the empirical CDF of dataset $x_1,x_2,...x_N$, and $x_i$ is the PIT value at $\tilde{\theta}_i$ of the $i$th sample. $F(x)$ is the CDF of the theoretical uniform distribution. The null hypothesis of the KS test is that the variable $x_i$ observes a uniform distribution. The $p$-value of the KS test is the probability to obtain a sample data distribution with $D$ larger than the measured $D_{\mathrm{obs}}$. 
We implement the KS test with the {\tt SciPy} package under the ``exact'' mode \citep{JSSv039i11}. It is difficult to write down the exact formula of the $p$-value in the KS test. To give some intuition, it can be approximated by \citep{Ivezic2014} 
\begin{equation}
\operatorname{Pr}\left(D>D_{\mathrm{obs}}\right)=\mathcal{Q}_{\mathrm{KS}}\left([\sqrt{N}+0.12+0.11 / \sqrt{N}] \,D_{\mathrm{obs}}\right)\,.
\end{equation}
Here, the survival function $Q_{\mathrm{KS}} $ is defined as   
\begin{equation}
\mathcal{Q}_{\mathrm{KS}}(u)=2 \sum_{m=1}^{\infty}(-1)^{m-1} \exp \left(-2 m^{2} u^{2}\right).
\end{equation}

The CvM test is based on the distance measure $D^*$ defined as 
\begin{equation}
D^* \equiv N \int_{-\infty}^{\infty}\left(F_{N}(x)-F(x)\right)^{2} \mathrm{~d} F(x)\,.
\label{eqn:CvM}
\end{equation}
The $p$-value of the CvM test is the probability to obtain a sample data distribution with $D^*$ larger than the measured $D^*_{\mathrm{obs}}$, which is given \citep{https://doi.org/10.1111/j.2517-6161.1996.tb02077.x} in the {\tt SciPy} package. 

The KS test and the CvM test focus on different aspects of a distribution: while the KS test is sensitive to the median, the CvM test can capture the tails of a distribution. If the $p$-value is greater than some criterion, typically 0.01 or 0.05 \citep{Ivezic2014}, then the null hypothesis that the distribution is a uniform distribution can be accepted.

\subsection{Joint Posteriors Validation}

Since parameters are degenerate, validation of the marginal posterior can be biased. The tools in this subsection focus on the validation of posteriors in the joint parameter space. 

\subsubsection{Probabilistic Copula Calibration}

The probabilistic copula calibration \citep{10.1214/14-EJS964,mucesh2021machine} is an extension of the probabilistic calibration. Consider a joint parameter distribution $f(\boldsymbol{\theta})$, where $\boldsymbol{\theta}$ is a vector of parameters. For example, $\boldsymbol{\theta} = \{ \log _ { 10 } \left( T_ { \rm vir } \right),\,\mathrm{log_{10}\zeta}\}$ in this paper, and $f(\boldsymbol{\theta})$ is the multi-dimensional probability distribution function that is generated in the MCMC chain given the observed data. 

We define the Kendall distribution function 
\begin{equation}
\mathcal{K}_{H}(w)\equiv \operatorname{Pr}\{H(\boldsymbol{\theta}) \leq w\} \quad \text { for } \quad w \in[0,1]\,,
\end{equation}
where $H(\boldsymbol{\theta})$ is the CDF of the distribution $f(\boldsymbol{\theta})$, and ``Pr'' means the probability. 
Thus the Kendall distribution function can be interpreted as the CDF of the CDF. 

We define the copula probability integral transformation (copPIT) as the Kendall distribution function evaluated at the CDF of the true value $H(\tilde{\boldsymbol{\theta}})$, i.e.\ 
\begin{equation}
{\rm copPIT} \equiv \mathcal{K}_{H}(H(\tilde{\boldsymbol{\theta}})) = \operatorname{Pr}\{H(\boldsymbol{\theta}) \leq H(\tilde{\boldsymbol{\theta}})\}.
\end{equation}

The joint posteriors are probabilistically copula calibrated if the copPIT is uniformly distributed. In the QQ plot, wherein the quantile of the copPIT distribution from the data is compared with that from a uniform distribution of copPIT, then this QQ plot falls on the diagonal line if the copPIT distribution is exactly uniform.

\subsubsection{Kendall Calibration}

As an extension of the marginal calibration, the Kendall calibration \citep{10.1214/14-EJS964,mucesh2021machine} compares the average Kendall distribution function with the empirical CDF. 
The average Kendall distribution function is defined as 
\begin{equation}
\hat{\mathcal{K}}_{H_{I}}(w)\equiv\frac{1}{N} \sum_{i=1}^{N} \mathcal{K}_{H_{i}}(w)\,,
\end{equation}
where $\mathcal{K}_{H_{i}}(w)$ is the Kendall distribution function of the CDF $H_{i}(\boldsymbol{\theta})$ for the $i$th sample.
The empirical CDF of the joint CDFs evaluated at the true values $\tilde{\boldsymbol{\theta}}$ is defined as 
\begin{equation}
\tilde{J}_{I}(w) \equiv \frac{1}{N} \sum_{i=1}^{N} \mathbbm{1}\left\{H_{i}\left(\tilde{\boldsymbol{\theta}}_{i}\right) \leq w\right\}\,.
\end{equation}

Kendall calibration tests whether $\hat{\mathcal{K}}_{H_{I}}(w) - \tilde{J}_{I}(w)$ is small for any value $0\le w\le 1$. In other words, Kendall calibration probes how well the inferred distribution agrees with the real distribution on average.

\subsubsection{Probabilistic HPD Calibration}

The highest probability density (HPD) is defined as \citep{harrison2015validation} 
\begin{equation}
{\rm HPD}\,(\boldsymbol{\tilde{\theta}}) \equiv \int_{f(\boldsymbol{\theta}) \geq f(\boldsymbol{\tilde{\theta}})} f(\boldsymbol{\theta}) \mathrm{d}^{n} \boldsymbol{\theta}\,,
\end{equation}
where $\mathrm{d}^{n} \boldsymbol{\theta}$ is a volume element in the multi-dimensional parameter space. 

The HPD describes the plausibility of $\boldsymbol{\tilde{\theta}}$ under the distribution $f(\boldsymbol{\theta})$. Small value indicates high plausibility. The joint posteriors are probabilistically HPD calibrated, if the HPD is uniformly
distributed. 

Note that for the PIT, copPIT, and HPD, the uniformity of their distributions is a necessary, but not sufficient, condition for the accurate posteriors \citep{gneiting2007probabilistic,10.1214/14-EJS964,harrison2015validation,2021arXiv210210473Z}.

\subsubsection{Hypothesis Tests}

The data $x$, or the PIT value, in Equations~(\ref{eqn:KS}) and (\ref{eqn:CvM}) can be replaced by the copPIT value and the HPD value, in the KS test and CvM test, respectively. 

\begin{figure*}
\centering
    \includegraphics[width=\textwidth]{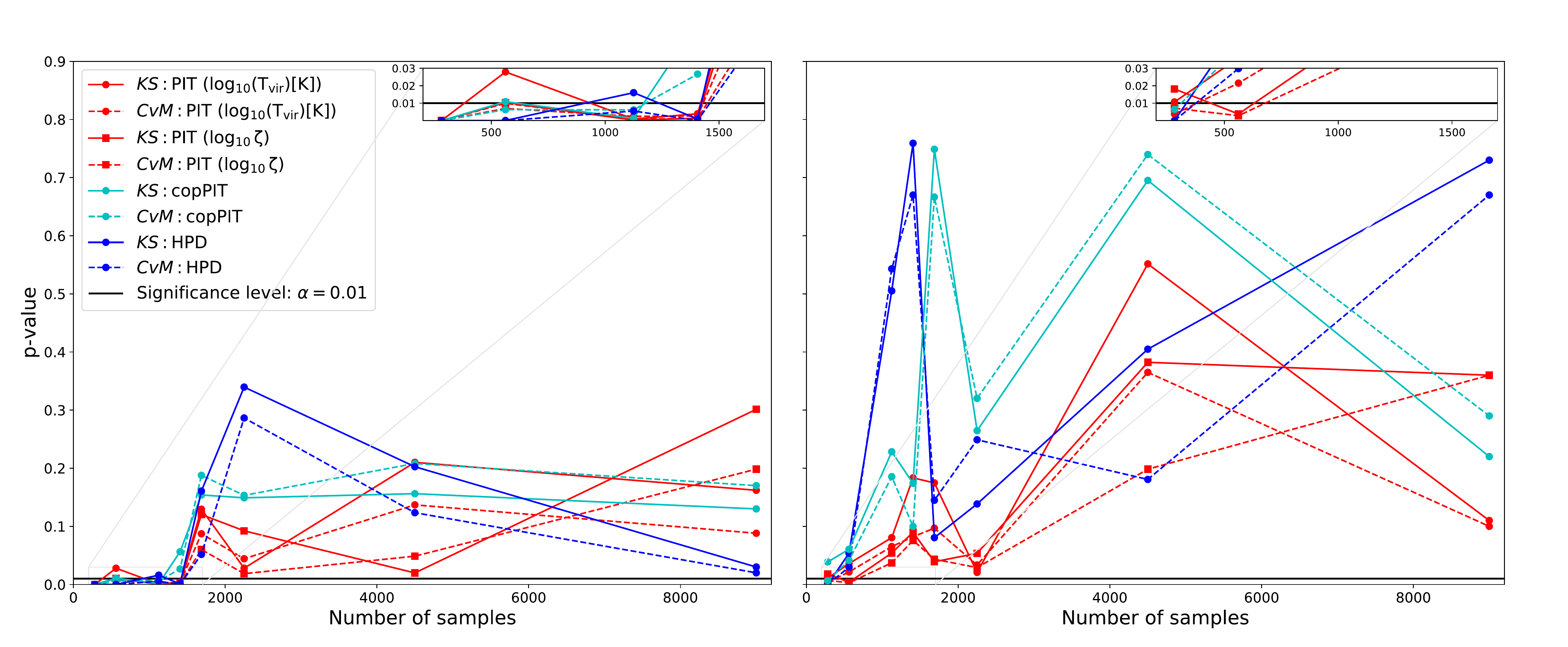}
    \caption{The effect of the training sample size on the posterior inference. We show the p-values of the KS test (solid line) and CvM test (dashed line) for the null hypothesis that a statistic is of a uniform distribution, as a function of the training sample size. The tests are performed over 300 testing samples with the mock observations applied with the thermal noise of HERA (left) and SKA (right). The statistical metrics include the PIT for $\log _ { 10 } \left( T_ { \rm vir } \right)$ (red dots) and  $\mathrm{log_{10}\zeta}$ (red squares), the copPIT (cyan dots) and HPD (blue dots). The threshold of the p-value above which the null hypothesis is accepted is marked (black solid line).}
    \label{fig:size}
\end{figure*}

\section{Effect of the Sample Size} 
\label{sec: size}

In mock observations with thermal noise and foreground cut, 9000 samples were employed for training in this paper. Nevertheless, this size of database might be more than necessary. In this section, we investigate the effect of the training sample size on the posterior inference.

The p-value of the null hypothesis that a statistic is of a uniform distribution is a good indicator for validating the inferred posteriors. We adopt the criterion that the significance level is above $0.01$ for accepting the null hypothesis. The p-value is typically affected by the sample size, as well as the complexity of the individual NDE and the ensembles of multiple NDEs. With the chosen NDE architecture and the ensemble of five such NDEs, we use the bisection method and decrease the sample size for training the NDEs to find out the minimum size of training samples that meets the criterion of accepting the null hypothesis.

In Fig.~\ref{fig:size}, we plot the p-values for the null hypotheses of the statistics as a function of the training sample size. For the mock HERA (SKA) observations, the p-values for all statistics are above $0.01$ when the sample size is larger than about 1500 (1000). This test indicates that the minimum sample size for training can be about 1500, in general. However, since the hypothesis tests are only necessary, not sufficient, conditions for accurate posteriors, a training dataset of only 1500 samples may not be optimal. Also, our estimate of the minimum sample size is only based on a two-parameter space of reionization model. The scaling law in a higher dimensional parameter space is beyond the scope of this paper.

\bibliographystyle{aasjournal}
\bibliography{Ref}{}



\end{document}